\pdfoutput=1
\documentclass[12pt,a4paper]{article}

\usepackage{ifthen} 
\newboolean{pdflatex}
\setboolean{pdflatex}{true} 

\newboolean{articletitles}
\setboolean{articletitles}{true} 

\newboolean{uprightparticles}
\setboolean{uprightparticles}{true} 

\newboolean{inbibliography}
\setboolean{inbibliography}{false} 

\def\paperauthors{LHCb collaboration} 
\def\paperasciititle{Observation of an exotic narrow 
doubly charmed tetraquark}
\def\papertitle{Observation of an~exotic narrow 
doubly charmed tetraquark} 
\def\paperkeywords{{High Energy Physics}, {LHCb}} 
\def\papercopyright{\the\year\ CERN 
for the benefit of the LHCb collaboration} 
\def\paperlicence{CC BY 4.0 licence}
\def\paperlicenceurl{https://creativecommons.org/licenses/by/4.0/}

\usepackage{lscape}
\usepackage{comment}
\usepackage{graphpap} 
\usepackage{multirow} 
\usepackage{rotating} 
\usepackage{mathrsfs}
\usepackage{mathtools} 
\usepackage{dashrule}
\usepackage{tikz}
\usepackage{enumerate}
\usepackage{enumitem}
\usetikzlibrary{patterns}
\usepackage{nicefrac} 
\usepackage{afterpage}
\usepackage{xcolor}
\usepackage[toc,page]{appendix}
\usepackage[normalem]{ulem} 
\usepackage{wasysym}

\usepackage{xpatch}

\newcounter{mybibstartvalue}
\xpatchcmd{\thebibliography}{%
  \usecounter{enumiv}%
}{%
  \usecounter{enumiv}%
  \setcounter{enumiv}{\value{mybibstartvalue}}%
}{}{}

\usepackage[top=1in, bottom=1.25in, left=1in, right=1in]{geometry}

\usepackage{microtype}
\usepackage{lineno}  
\usepackage{xspace} 
\usepackage{caption} 

\usepackage{graphicx}  
\usepackage{color}
\usepackage{colortbl}
\graphicspath{{./figs/}} 

\usepackage{amsmath} 
\usepackage{amssymb}
\usepackage{amsfonts}
\usepackage{upgreek} 

\newcommand*\patchAmsMathEnvironmentForLineno[1]{%
\expandafter\let\csname old#1\expandafter\endcsname\csname #1\endcsname
\expandafter\let\csname oldend#1\expandafter\endcsname\csname
end#1\endcsname
 \renewenvironment{#1}%
   {\linenomath\csname old#1\endcsname}%
   {\csname oldend#1\endcsname\endlinenomath}%
}
\newcommand*\patchBothAmsMathEnvironmentsForLineno[1]{%
  \patchAmsMathEnvironmentForLineno{#1}%
  \patchAmsMathEnvironmentForLineno{#1*}%
}
\AtBeginDocument{%
\patchBothAmsMathEnvironmentsForLineno{equation}%
\patchBothAmsMathEnvironmentsForLineno{align}%
\patchBothAmsMathEnvironmentsForLineno{flalign}%
\patchBothAmsMathEnvironmentsForLineno{alignat}%
\patchBothAmsMathEnvironmentsForLineno{gather}%
\patchBothAmsMathEnvironmentsForLineno{multline}%
\patchBothAmsMathEnvironmentsForLineno{eqnarray}%
}

\usepackage{hyperxmp}

\usepackage[pdftex,
            pdfauthor={\paperauthors},
            pdftitle={\paperasciititle},
            pdfkeywords={\paperkeywords},
            pdfcopyright={Copyright (C) \papercopyright},
            pdflicenseurl={\paperlicenceurl}]{hyperref}

\usepackage[colorinlistoftodos,textsize=scriptsize]{todonotes}

\usepackage[bottom,flushmargin,hang,multiple]{footmisc}

\usepackage[all]{hypcap} 

\usepackage{xspace} 
\usepackage{upgreek}

\def\lhcb   {\mbox{LHCb}\xspace}

\def\MagUp {\mbox{\em Mag\kern -0.05em Up}\xspace}

\ifthenelse{\boolean{uprightparticles}}%
{
 
 \def\Pgamma      {\ensuremath{\upgamma}\xspace}

 \def\Pmu         {\ensuremath{\upmu}\xspace}

 \def\Ppi         {\ensuremath{\uppi}\xspace}

 \def\Ppsi        {\ensuremath{\uppsi}\xspace}

 \def\PDelta      {\ensuremath{\Delta}\xspace}                 
 \def\PXi         {\ensuremath{\Xi}\xspace}                 
 \def\PLambda     {\ensuremath{\Lambda}\xspace}                 
 \def\PSigma      {\ensuremath{\Sigma}\xspace}                 
 \def\POmega      {\ensuremath{\Omega}\xspace}                 
 \def\PUpsilon    {\ensuremath{\Upsilon}\xspace}                 

 \def\PA      {\ensuremath{\mathrm{A}}\xspace}                 
 \def\PB      {\ensuremath{\mathrm{B}}\xspace}                 
                  
 \def\PD      {\ensuremath{\mathrm{D}}\xspace}

 \def\PJ      {\ensuremath{\mathrm{J}}\xspace}                 
 \def\PK      {\ensuremath{\mathrm{K}}\xspace}

 \def\PQ      {\ensuremath{\mathrm{Q}}\xspace}

 \def\PT      {\ensuremath{\mathrm{T}}\xspace}

 \def\Pb      {\ensuremath{\mathrm{b}}\xspace}                 
 \def\Pc      {\ensuremath{\mathrm{c}}\xspace}                 
 \def\Pd      {\ensuremath{\mathrm{d}}\xspace}

 \def\Pi      {\ensuremath{\mathrm{i}}\xspace}

 \def\Pp      {\ensuremath{\mathrm{p}}\xspace}                 
 \def\Pq      {\ensuremath{\mathrm{q}}\xspace}                 
                  
 \def\Ps      {\ensuremath{\mathrm{s}}\xspace}                 
                  
 \def\Pu      {\ensuremath{\mathrm{u}}\xspace}

 \def\thebaroffset{0.0em}
}
{
 
 \def\Pgamma      {\ensuremath{\gamma}\xspace}

 \def\Pmu         {\ensuremath{\mu}\xspace}

 \def\Ppi         {\ensuremath{\pi}\xspace}

 \def\Ppsi        {\ensuremath{\psi}\xspace}                 
                  
 \mathchardef\PDelta="7101
 \mathchardef\PXi="7104
 \mathchardef\PLambda="7103
 \mathchardef\PSigma="7106
 \mathchardef\POmega="710A
 \mathchardef\PUpsilon="7107
 \def\PA      {\ensuremath{A}\xspace}                 
 \def\PB      {\ensuremath{B}\xspace}                 
                  
 \def\PD      {\ensuremath{D}\xspace}

 \def\PJ      {\ensuremath{J}\xspace}                 
 \def\PK      {\ensuremath{K}\xspace}

 \def\PQ      {\ensuremath{Q}\xspace}

 \def\PT      {\ensuremath{T}\xspace}

 \def\Pb      {\ensuremath{b}\xspace}                 
 \def\Pc      {\ensuremath{c}\xspace}                 
 \def\Pd      {\ensuremath{d}\xspace}

 \def\Pi      {\ensuremath{i}\xspace}

 \def\Pp      {\ensuremath{p}\xspace}                 
 \def\Pq      {\ensuremath{q}\xspace}                 
                  
 \def\Ps      {\ensuremath{s}\xspace}                 
                  
 \def\Pu      {\ensuremath{u}\xspace}

 \def\thebaroffset{0.18em}
}
\newcommand{\offsetoverline}[2][\thebaroffset]{\kern #1\overline{\kern -#1 #2}}%

\makeatletter
\ifcase \@ptsize \relax
  \newcommand{\miniscule}{\@setfontsize\miniscule{4}{5}}
\or
  \newcommand{\miniscule}{\@setfontsize\miniscule{5}{6}}
\or
  \newcommand{\miniscule}{\@setfontsize\miniscule{5}{6}}
\fi
\makeatother

\DeclareRobustCommand{\optbar}[1]{\shortstack{{\miniscule (\rule[.5ex]{1.25em}{.18mm})}
  \\ [-.7ex] $#1$}}

\def\mumu       {{\ensuremath{\Pmu^+\Pmu^-}}\xspace}

\def\g      {{\ensuremath{\Pgamma}}\xspace}

\def\quark     {{\ensuremath{\Pq}}\xspace}
\def\quarkbar  {{\ensuremath{\overline \quark}}\xspace}

\def\uquark    {{\ensuremath{\Pu}}\xspace}
\def\uquarkbar {{\ensuremath{\overline \uquark}}\xspace}

\def\dquark    {{\ensuremath{\Pd}}\xspace}
\def\dquarkbar {{\ensuremath{\overline \dquark}}\xspace}

\def\squark    {{\ensuremath{\Ps}}\xspace}

\def\cquark    {{\ensuremath{\Pc}}\xspace}

\def\bquark    {{\ensuremath{\Pb}}\xspace}

\def\pion   {{\ensuremath{\Ppi}}\xspace}
\def\piz    {{\ensuremath{\pion^0}}\xspace}
\def\pip    {{\ensuremath{\pion^+}}\xspace}
\def\pim    {{\ensuremath{\pion^-}}\xspace}

\def\kaon    {{\ensuremath{\PK}}\xspace}

\def\KorKbar {\kern \thebaroffset\optbar{\kern -\thebaroffset \PK}{}\xspace}

\def\Kp      {{\ensuremath{\kaon^+}}\xspace}
\def\Km      {{\ensuremath{\kaon^-}}\xspace}

\def\KS      {{\ensuremath{\kaon^0_{\mathrm{S}}}}\xspace}

\def\Dbar    {{\ensuremath{\offsetoverline{\PD}}}\xspace}
\def\D       {{\ensuremath{\PD}}\xspace}

\def\DorDbar {\kern \thebaroffset\optbar{\kern -\thebaroffset \PD}\xspace}
\def\Dz      {{\ensuremath{\D^0}}\xspace}
\def\Dzb     {{\ensuremath{\Dbar{}^0}}\xspace}
\def\Dp      {{\ensuremath{\D^+}}\xspace}
\def\Dm      {{\ensuremath{\D^-}}\xspace}

\def\DpDm    {\ensuremath{\Dp {\kern -0.16em \Dm}}\xspace}
\def\Dstar   {{\ensuremath{\D^*}}\xspace}

\def\Dstarz  {{\ensuremath{\D^{*0}}}\xspace}

\def\Dstarp  {{\ensuremath{\D^{*+}}}\xspace}

\def\B       {{\ensuremath{\PB}}\xspace}
\def\Bbar    {{\ensuremath{\offsetoverline{\PB}}}\xspace}

\def\BorBbar {\kern \thebaroffset\optbar{\kern -\thebaroffset \PB}\xspace}

\def\Bzb     {{\ensuremath{\Bbar{}^0}}\xspace}
\def\Bd      {{\ensuremath{\B^0}}\xspace}

\def\BdorBdbar {\kern \thebaroffset\optbar{\kern -\thebaroffset \Bd}\xspace}
\def\Bu      {{\ensuremath{\B^+}}\xspace}
\def\Bub     {{\ensuremath{\B^-}}\xspace}

\def\Bm      {{\ensuremath{\Bub}}\xspace}

\def\Bs      {{\ensuremath{\B^0_\squark}}\xspace}

\def\BsorBsbar {\kern \thebaroffset\optbar{\kern -\thebaroffset \Bs}\xspace}

\def\jpsi     {{\ensuremath{{\PJ\mskip -3mu/\mskip -2mu\Ppsi}}}\xspace}

\def\Y#1S{\ensuremath{\PUpsilon{(#1S)}}\xspace}

\def\proton      {{\ensuremath{\Pp}}\xspace}

\def\LorLbar     {\kern \thebaroffset\optbar{\kern -\thebaroffset \PLambda}\xspace}

\def\Xires       {{\ensuremath{\PXi}}\xspace}

\def\Xiccpp      {{\ensuremath{\Xires^{++}_{\cquark\cquark}}}\xspace}

\newcommand{\decay}[2]{\ensuremath{#1\!\to #2}\xspace} 

\def\to                 {\ensuremath{\rightarrow}\xspace}

\def\AT#1     {\ensuremath{A_{\mathrm{T}}^{#1}}\xspace}           

\def\C#1      {\ensuremath{\mathcal{C}_{#1}}\xspace}                       
\def\Cp#1     {\ensuremath{\mathcal{C}_{#1}^{'}}\xspace}                    
\def\Ceff#1   {\ensuremath{\mathcal{C}_{#1}^{\mathrm{(eff)}}}\xspace}        
\def\Cpeff#1  {\ensuremath{\mathcal{C}_{#1}^{'\mathrm{(eff)}}}\xspace}       
\def\Ope#1    {\ensuremath{\mathcal{O}_{#1}}\xspace}                       
\def\Opep#1   {\ensuremath{\mathcal{O}_{#1}^{'}}\xspace}                    

\newcommand{\nospaceunit}[1]{\ensuremath{\text{#1}}}       
\newcommand{\aunit}[1]{\ensuremath{\text{\,#1}}}       

\newcommand{\tev}{\aunit{Te\kern -0.1em V}\xspace}
\newcommand{\gev}{\aunit{Ge\kern -0.1em V}\xspace}
\newcommand{\mev}{\aunit{Me\kern -0.1em V}\xspace}
\newcommand{\kev}{\aunit{ke\kern -0.1em V}\xspace}
\newcommand{\ev}{\aunit{e\kern -0.1em V}\xspace}
 
\newcommand{\mevc}{\ensuremath{\aunit{Me\kern -0.1em V\!/}c}\xspace}
\newcommand{\gevc}{\ensuremath{\aunit{Ge\kern -0.1em V\!/}c}\xspace}
\newcommand{\mevcc}{\ensuremath{\aunit{Me\kern -0.1em V\!/}c^2}\xspace}
\newcommand{\gevcc}{\ensuremath{\aunit{Ge\kern -0.1em V\!/}c^2}\xspace}

\def\mum  {\ensuremath{\,\upmu\nospaceunit{m}}\xspace}

\def\fb   {\ensuremath{\aunit{fb}}\xspace}
\def\invfb   {\ensuremath{\fb^{-1}}\xspace}

\def\gsim{{~\raise.15em\hbox{$>$}\kern-.85em
          \lower.35em\hbox{$\sim$}~}\xspace}
\def\lsim{{~\raise.15em\hbox{$<$}\kern-.85em
          \lower.35em\hbox{$\sim$}~}\xspace}

\def\sPlot{\mbox{\em sPlot}\xspace}

\def\pt         {\ensuremath{p_{\mathrm{T}}}\xspace}

\def\ptot       {\ensuremath{p}\xspace}

\def\evtgen     {\mbox{\textsc{EvtGen}}\xspace}

\def\geant      {\mbox{\textsc{Geant4}}\xspace}

\def\pythia     {\mbox{\textsc{Pythia}}\xspace}

\def\tell1  {TELL1\xspace}
\def\ukl1   {UKL1\xspace}

\newcommand{\ie}{\mbox{\itshape i.e.}\xspace}

\usepackage{cite} 
\usepackage{mciteplus}

\usepackage{longtable} 

\newboolean{wordcount}
\setboolean{wordcount}{false} 

\newboolean{nature}
\setboolean{nature}{true} 

\newboolean{preliminary}
\setboolean{preliminary}{false}

\makeatletter
\g@addto@macro\bfseries{\boldmath}
\makeatother

\def\Tcc {{{\ensuremath{\PT_{\cquark\cquark}^+}}}\xspace}

\newcommand{\kevc}{\ensuremath{\aunit{ke\kern -0.1em V\!/}c}\xspace}
\newcommand{\kevcc}{\ensuremath{\aunit{ke\kern -0.1em V\!/}c^2}\xspace}

\usepackage{scalerel}

\def\XXint#1#2#3{{\setbox0=\hbox{$#1{#2#3}{\int}$}
     \vcenter{\hbox{$#2#3$}}\kern-.5\wd0}}

\begin{document}


\renewcommand{\thefootnote}{\fnsymbol{footnote}}
\setcounter{footnote}{1}

\ifthenelse{\boolean{wordcount}}{}{


\begin{titlepage}
\pagenumbering{roman}

\vspace*{-1.5cm}
\centerline{\large EUROPEAN ORGANIZATION FOR NUCLEAR RESEARCH (CERN)}
\vspace*{1.5cm}
\noindent
\begin{tabular*}{\linewidth}{lc@{\extracolsep{\fill}}r@{\extracolsep{0pt}}}
\ifthenelse{\boolean{pdflatex}}
{\vspace*{-1.5cm}\mbox{\!\!\!\includegraphics[width=.14\textwidth]{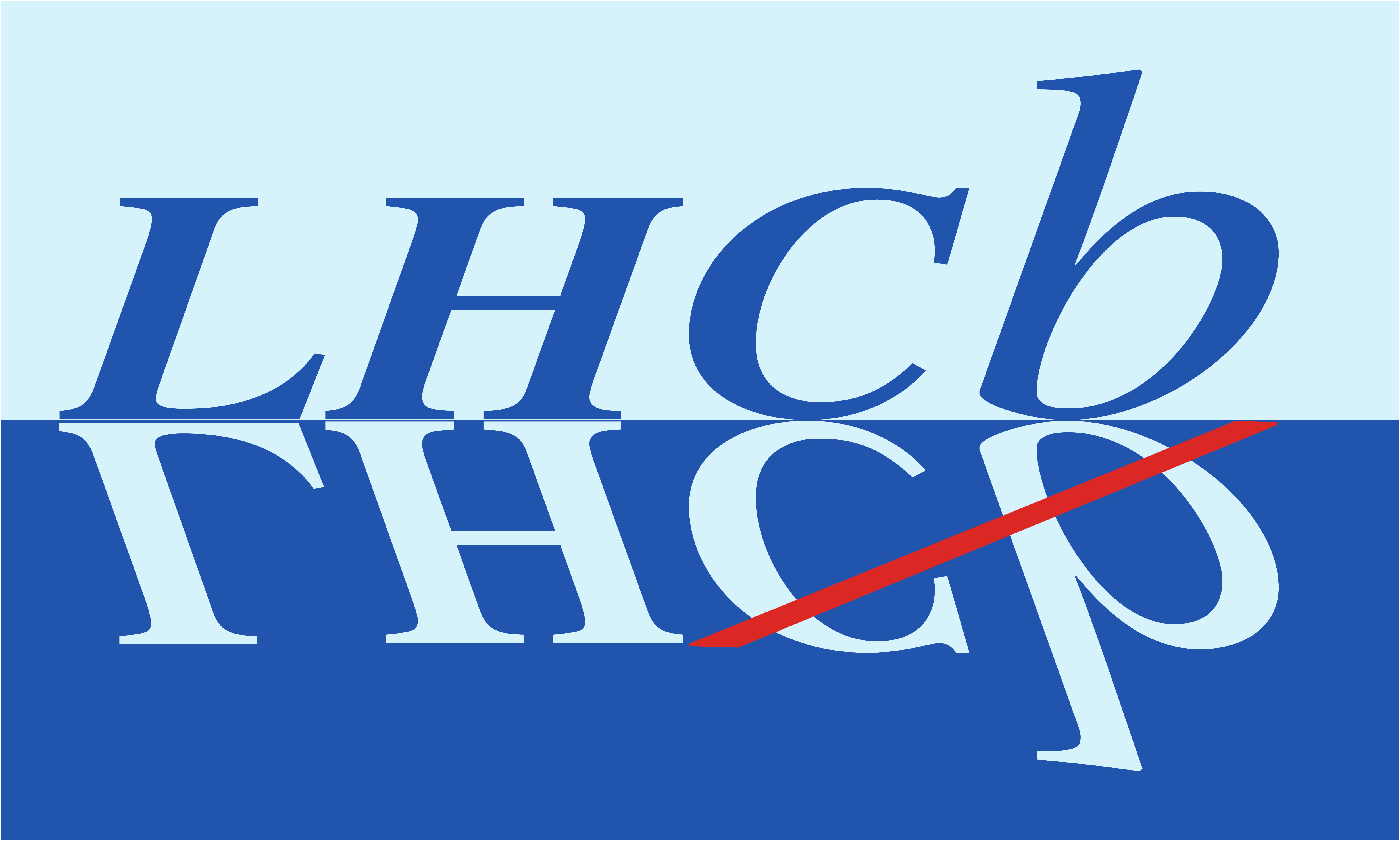}} & &}%
{\vspace*{-1.2cm}\mbox{\!\!\!\includegraphics[width=.12\textwidth]{lhcb-logo.eps}} & &}%
\\
 & & CERN-EP-2021-165 \\  
 & & LHCb-PAPER-2021-031 \\  
  &  &  September 2, 2021
\end{tabular*}

\vspace*{2.0cm}

{\normalfont\bfseries\boldmath\huge
\begin{center}
  \papertitle 
\end{center}
}

\vspace*{1.5cm}

\begin{center}
\paperauthors\footnote{Authors are listed at the end of this Letter.}
\end{center}

\vspace{\fill}

\begin{abstract}
\noindent 
Conventional hadronic matter consists 
of baryons and mesons made of three quarks 
and quark\nobreakdash-antiquark pairs, 
respectively~\cite{GellMann:1964nj,
Zweig:352337}. 
Here, we~report the~observation of a~hadronic state 
containing four quarks with the~LHCb experiment 
at the~Large Hadron Collider. 
This so\nobreakdash-called tetraquark contains 
two charm quarks, a~\uquarkbar and a~\dquarkbar~quark. 
This~exotic state has a~mass of approximately 
3875\mevcc and manifests itself as a~narrow peak 
in the~mass spectrum of \Dz\Dz\pip~mesons just below 
the~$\Dstarp\Dz$~mass threshold. 
The~near\nobreakdash-threshold mass together with 
the~narrow width reveals the~resonance nature of the~state.
\end{abstract}

\vspace*{5.0cm}

\begin{center}
Published in  \href{https://doi.org/10.1038/s41567-022-01614-y}{Nature Physics (2021)}
\end{center}

\vspace{\fill}

{\footnotesize 
\centerline{\copyright~\papercopyright. \href{\paperlicenceurl}{\paperlicence}.}}

\end{titlepage}


\newpage
\setcounter{page}{2}
\mbox{~}

}


\renewcommand{\thefootnote}{\arabic{footnote}}
\setcounter{footnote}{0}

\cleardoublepage


\pagestyle{plain} 
\setcounter{page}{1}
\pagenumbering{arabic}



Quantum chromodynamics\,(QCD), 
the~theory of the~strong force, 
describes 
interactions of coloured quarks and gluons
and the~formation of hadronic matter, 
the~so-called mesons and baryons. 
While QCD makes precise 
predictions at high energies, 
the theory has difficulties
describing 
the~interactions of  quarks
in hadrons 
from first principles 
due to the~highly 
non\nobreakdash-perturbative 
regime at the~corresponding energy scale.
Hence, the~field of hadron spectroscopy is
driven by
experimental discoveries
that sometimes are unexpected, 
which could lead to changes 
in the~research landscape. 
Along with 
conventional mesons and baryons, 
made of 
a~quark-antiquark 
pair\,(\mbox{$\quark_1\quarkbar_2$})
and  three 
quarks\,(\mbox{$\quark_1\quark_2\quark_3$}),
respectively, 
particles with 
an~alternative quark content,
known as exotic states, 
have been actively discussed 
since the~birth of the~constituent quark 
model~\cite{GellMann:1964nj,
Zweig:352337,Zweig:570209, 
Jaffe:1976ig,
Jaffe:1976ih,
Rossi:1977cy,
Jaffe:1977cv, 
Lipkin:1987sk}.
The~discussion has been revived 
by recent observations of 
numerous 
tetraquark $\quark_1\quark_2\quarkbar_3\quarkbar_4$ 
and pentaquark 
$\quark_1\quark_2\quark_3\quark_4\quarkbar_5$ 
candidates~\cite{
Choi:2003ue,
 Choi:2007wga,
 Belle:2008qeq,
  Aaltonen:2009tz,
 Belle:2011aa, 
  Ablikim:2013mio,
  BESIII:2013ouc,
  BESIII:2013mhi,
  Abazov:2013xda,
  Chatrchyan:2013dma,
   Chilikin:2013tch,
   Belle:2013yex, 
   Chilikin:2014bkk, 
  LHCb-PAPER-2014-014,
   LHCb-PAPER-2015-029, 
  LHCb-PAPER-2015-038,
  LHCb-PAPER-2016-018, 
  LHCb-PAPER-2016-019, 
  LHCb-PAPER-2018-034,
  LHCb-PAPER-2019-014,
  LHCb-PAPER-2020-011, 
  LHCb-PAPER-2020-024,
  LHCb-PAPER-2020-025, 
  LHCb-PAPER-2020-035,
  LHCb-PAPER-2020-039,
  LHCb-PAPER-2020-044,
   Ablikim_2021,
  LHCb-PAPER-2021-018}. 
Due~to the~closeness of their masses 
to known particle\nobreakdash-pair 
thresholds~\cite{
Brambilla:2019esw,
PDG2021}, 
many of these 
states are likely to be 
hadronic molecules~\cite{Oset:2019upy,
Richard_2016,Guo:2017jvc, 
MartinezTorres:2020hus}
where colour\nobreakdash-singlet hadrons are bound by 
 residual nuclear forces
similar to the~electromagnetic 
van der Waals 
forces attracting 
electrically neutral atoms and molecules.
An~ordinary 
example of 
a~hadronic molecule is the~deuteron 
formed by a~proton and a~neutron.
On~the~other hand, an~interpretation of exotic states
as compact multiquark structures 
is also possible~\cite{Ali:2019roi}.

All exotic hadrons observed so far 
predominantly
decay via 
the~strong interaction and 
their decay widths vary from a~few to a~few hundred~\mev.
A~discovery of a~long\nobreakdash-lived exotic 
state, stable 
with respect to the~strong interaction,
would be intriguing.
A~hadron with two heavy quarks~$\PQ$ and 
two light antiquarks~$\quarkbar$, \ie
\mbox{$\PQ_1\PQ_2\quarkbar_1\quarkbar_2$}, 
is a~prime candidate to form such 
a~state~\cite{Ader:1981db,
Ballot:1983iv,
Zouzou:1986qh,
Lipkin:1986dw,
Heller:1986bt,
Manohar:1992nd}.
In~the~limit of a~large heavy-quark mass
the~two heavy 
quarks $\PQ_1\PQ_2$ form a~point\nobreakdash-like 
heavy colour\nobreakdash-antitriplet object,
that behaves similarly to an~antiquark,  
and the~corresponding state should be bound.
It is expected that the~$\bquark$ quark 
is heavy enough to sustain
the~existence of a~stable 
\mbox{$\bquark\bquark\uquarkbar\dquarkbar$}~state 
with its binding energy of
about $200\mev$
with
respect to 
the~sum of masses of the~pseudoscalar, 
$\Bm$ or $\Bzb$,
and vector, 
$\B^{*-}$ or $\Bbar{}^{*0}$,
beauty mesons
that defines the~minimal mass 
for the~strong decay to be allowed.
In~the~case of 
the~\mbox{$\bquark\cquark\uquarkbar\dquarkbar$}
and~\mbox{$\cquark\cquark\uquarkbar\dquarkbar$}~systems,
there is currently 
no consensus  as to 
whether such states exist and are narrow 
enough to be detected experimentally.
The~similarity of 
the~\cquark\cquark\uquarkbar\dquarkbar~tetraquark state 
and the~\Xiccpp~baryon
containing two \cquark~quarks 
and a~\uquark~quark, 
leads to a~relationship between 
the~properties of the~two states.
In~particular, the~measured mass of 
the~\Xiccpp~baryon
with quark content $\cquark\cquark\uquark$~\cite{
LHCb-PAPER-2017-018,
LHCb-PAPER-2018-026,
LHCb-PAPER-2019-037}
implies that the~mass 
of the~$\cquark\cquark\uquarkbar\dquarkbar$
tetraquark is close to the
sum of masses of \Dz and \Dstarp~mesons 
with quark content of 
$\cquark\uquarkbar$ and $\cquark\dquarkbar$,
respectively, 
as suggested in Ref.~\cite{Karliner:2017qjm}.
Theoretical predictions for 
the~mass of  
the~$\mbox{\cquark\cquark\uquarkbar\dquarkbar}$~ground~state
with 
spin\nobreakdash-parity quantum 
numbers $\mathrm{J^P=1^+}$ and 
isospin $\mathrm{I}=0$,
denoted hereafter as \Tcc, 
relative to the~$\Dstarp\Dz$~mass threshold
\ifthenelse{\boolean{wordcount}}{}{
\begin{equation}\label{eq:delta_prime}
 \updelta m  \equiv 
 m_{\Tcc} - \left( m_{\Dstarp} + m_{\Dz}\right) \,,
\end{equation}
}
lie in the~range 
\mbox{$-300< \updelta m < 300\mevcc$}~\cite{
Carlson:1987hh,
SilvestreBrac:1993ss,
Semay:1994ht,
Moinester:1995fk,
Pepin:1996id,
GELMAN2003296,
Vijande:2003ki,
Janc:2004qn,
Navarra:2007yw,
Vijande,
Ebert:2007rn,
Lee:2009rt,
Yang2009,
Li:2012ss,
Feng:2013kea,
Luo:2017eub,
Karliner:2017qjm,
Eichten:2017ffp,
Wang:2017uld,
Park:2018wjk,
Junnarkar:2018twb,
Deng:2018kly,
Liu:2019stu,
Maiani:2019lpu,
Yang:2019itm,
Tan:2020ldi,
Lu:2020rog,
Braaten:2020nwp,
Gao:2020ogo,
Cheng:2020wxa,
Noh:2021lqs,
Faustov:2021hjs},
where $m_{\Dstarp}$ and $m_{\Dz}$ denote 
the~known masses of the~$\Dstarp$ and $\Dz$~mesons~\cite{PDG2021}.
Lattice QCD calculations also do not provide 
a~definite conclusion on the~existence 
of the~\Tcc~state 
and its~binding energy~\cite{
Ikeda:2013vwa,
Cheung:2017tnt,
Francis:2018jyb,
Junnarkar:2018twb}.
The~observation of 
the~$\Xiccpp$ baryon~\cite{LHCb-PAPER-2017-018,
LHCb-PAPER-2018-026}
and of a~new exotic resonance decaying to
a~pair of \jpsi~mesons~\cite{LHCb-PAPER-2020-011}
by the \lhcb experiment,
motivates the~search for the~\Tcc state.

In this Letter, the~observation of a~narrow state 
in the~\Dz\Dz\pip~mass 
spectrum 
near the $\Dstarp\Dz$~mass threshold 
compatible with 
being a~$\Tcc$~tetraquark state is reported.
Throughout this Letter, 
charge conjugate decays are implied.
The~study is based on 
proton\nobreakdash-proton\,($\proton\proton$) 
collision  data collected
with the~LHCb detector 
at the~Large Hadron Collider\,(LHC)
at CERN 
at centre\nobreakdash-of\nobreakdash-mass 
energies of~7,~8 and~13\tev, 
corresponding to integrated
luminosity of~9\invfb. 
The~LHCb detector~\cite{LHCb-DP-2008-001,LHCb-DP-2014-002} 
is a~single-arm forward spectrometer covering 
the~pseudorapidity
range \mbox{$2 < \eta < 5$}, 
designed for the~study of 
particles containing 
\bquark\ or \cquark\ quarks and
is further described in Methods.
Pseudorapidity $\eta$ is defined 
as $-\log \left( \tan \tfrac{\theta}{2}\right) $,
where $\theta$~is a~polar angle of the track relative to
the~proton beam line.

\begin{figure}[t]
  \includegraphics[width=\textwidth]{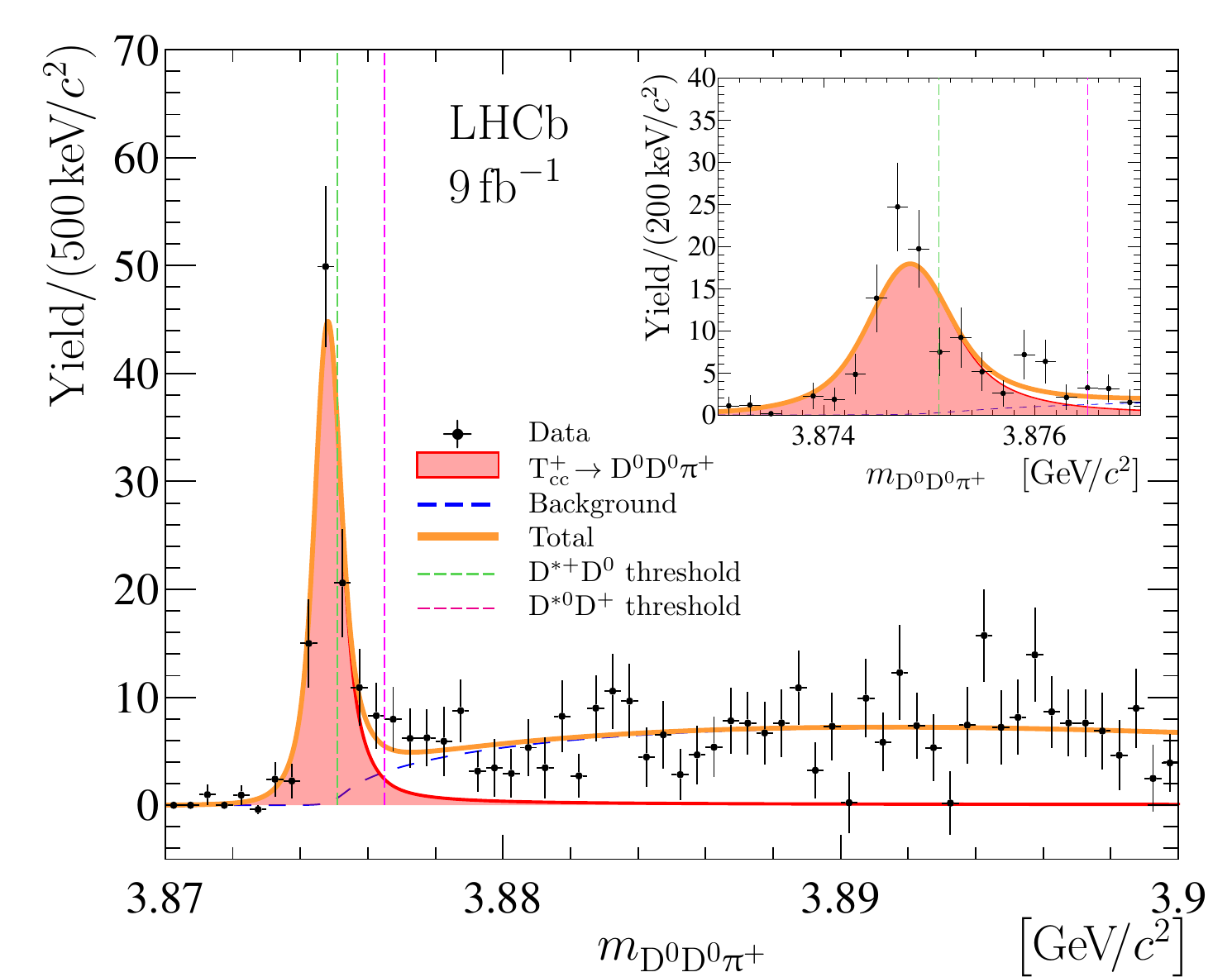}
  \caption { \small
 {\bf Distribution of \Dz\Dz\pip~mass}.
 Distribution of \Dz\Dz\pip~mass where 
 the~contribution of the~non\protect\nobreakdash-$\Dz$ 
 background has been statistically subtracted.
 The~result of the~fit 
 with the~two\nobreakdash-component function 
 described in the~text is overlaid.
 The~$\Dstarp\Dz$ and $\Dstarz\Dp$~thresholds 
 are indicated with the~vertical dashed lines. 
 The~horizontal bin width is indicated on 
 the~vertical axis legend.
 Inset shows a~zoomed signal region
 with fine binning scheme,  
 Uncertainties on the~data points are statistical
 only and represent one standard deviation, 
 calculated as a~sum in quadrature of the~assigned weights from 
 the~background\protect\nobreakdash-subtraction procedure.
 }
  \label{fig:DATA_BW0_fit}
\end{figure}

The~$\Dz\Dz\pip$ final state is reconstructed by 
selecting events with two \Dz~mesons and 
a~positively charged pion all produced 
at the~same $\proton\proton$ 
interaction point.
Both~\Dz~mesons are reconstructed 
in the~$\Dz\to\Km\pip$~decay channel. 
The~selection criteria are similar to those 
used in Ref.~\cite{LHCb-PAPER-2012-003}.
To~subtract background not originating 
from two \Dz~candidates
an~extended unbinned 
maximum\nobreakdash-likelihood fit to 
the~two\nobreakdash-dimensional distribution of
the~masses of the~two \Dz~candidates 
is performed. 
The corresponding procedure, 
together with the~selection criteria, is described in detail in Methods.
To~improve the~$\updelta m$  mass resolution and to make
the~determination insensitive 
to the~precision of the~\Dz~meson mass, 
the~mass of the~$\Dz\Dz\pip$~combinations is
calculated with the~mass of each $\Dz$~meson
constrained to the~known value~\cite{PDG2021}.
%
The~resulting 
$\Dz\Dz\pip$~mass distribution 
for~selected $\Dz\Dz\pip$~combinations
is shown in Fig.~\ref{fig:DATA_BW0_fit}.
A~narrow peak near 
the~$\Dstarp\Dz$~mass threshold is clearly visible. 

An~extended unbinned maximum\nobreakdash-likelihood 
fit to the~$\Dz\Dz\pip$~mass
distribution is performed using 
a~model consisting of signal and background components.
The~signal component
is described by the~convolution of 
the~detector resolution with 
a~resonant shape,
which is modelled by 
a~relativistic P\nobreakdash-wave 
two\nobreakdash-body Breit\nobreakdash--Wigner function 
modified by a~\mbox{Blatt}\nobreakdash--Weisskopf 
form factor 
with a~meson\nobreakdash-radius parameter of $3.5\gev^{-1}$. 
The~use of 
a~P\nobreakdash--wave resonance 
is motivated by 
the~expected $\mathrm{J^P}=1^+$ quantum 
numbers for the~\Tcc~state.
A~two\nobreakdash-body decay 
structure \mbox{$\decay{\Tcc}{\PA\PB}$} 
is assumed with $m_{\PA}=2m_{\Dz}$  and $m_{\PB}=m_{\pip}$,
where $m_{\pip}$~stands for 
the~known mass of the~\pip~meson.
Several alternative prescriptions are used for evaluation
of systematic uncertainties.
Despite its simplicity, 
the~model serves well to quantify 
the~existence of the~$\Tcc$ state and to measure 
its  properties, 
such as the~position and the~width
of the~resonance.
A~follow\nobreakdash-up study~\cite{LHCb-PAPER-2021-032}
investigates the~underlying nature of 
the~\Tcc~state,
expanding on 
the~modelling of the~signal shape
and determining its physical properties.
The~detector resolution  
is modelled by the~sum 
of two Gaussian functions 
with a~common mean, 
where the~additional parameters are 
taken from simulation\,(see Methods) with 
corrections applied~\cite{LHCb-PAPER-2020-008,
LHCb-PAPER-2020-009,
LHCb-PAPER-2020-035}. 
The~root mean square of the~resolution 
function is around 400\kevcc. 
A~study of the~$\Dz\pip$~mass distribution 
for  $\Dz\Dz\pip$~combinations 
in the~region above the~$\Dstarz\Dp$~mass threshold 
and below $3.9\gevcc$ shows 
that  approximately $90\%$ of 
all random $\Dz\Dz\pip$~combinations 
contain a~genuine $\Dstarp$~meson.
Based on this observation, the~background component
is parametrised by the~product of  
a~two\nobreakdash-body phase space 
function 
and  a~positive second\nobreakdash-order polynomial. 
The~resulting function is convolved 
with the~detector resolution.

The~fit results are shown in 
Fig.~\ref{fig:DATA_BW0_fit}, 
and the~parameters of interest, 
namely the~signal yield, $N$, 
the~mass parameter of the Breit\nobreakdash--Wigner
function relative to the~$\Dstarp\Dz$~mass threshold,  
\mbox{$\updelta m_{\mathrm{BW}}
\equiv m_{\mathrm{BW}}-(m_{\Dstarp}+m_{\Dz})$},
and the~width parameter, $\Gamma_{\mathrm{BW}}$,
are listed in Table~\ref{tab:DATA_BW0_fits}. 
The~statistical significance of the~observed 
\mbox{$\decay{\Tcc}{\Dz\Dz\pip}$}~signal is estimated 
using Wilks' theorem 
to be $22$~standard deviations.
The~fit suggests that~the~mass parameter of 
the~Breit\nobreakdash--Wigner shape  
is slightly below 
the~$\Dstarp\Dz$~mass threshold. 
The~statistical significance
of the~hypothesis $\updelta m_{\mathrm{BW}}<0$ is 
estimated to be 4.3~standard deviations.

\begin{table}[bt]
	\centering
	\caption{\small
        Parameters obtained
	obtained from the~fit to the~$\Dz\Dz\pip$~mass spectrum.
	Signal yield, $N$, 
	Breit\protect\nobreakdash--Wigner 
	mass relative to 
	$\Dstarp\Dz$~mass threshold, 
	$\updelta m_{\mathrm{BW}}$,
	and width, $\Gamma_{\mathrm{BW}}$,
        are listed.
	The~uncertainties are statistical only. 
	}
	\label{tab:DATA_BW0_fits}
	\vspace{2mm}
	\begin{tabular*}{0.40\textwidth}
	{@{\hspace{3mm}}l@{\extracolsep{\fill}}l@{\hspace{3mm}}}
	Parameter  & ~~~~Value 
   \\[1mm]
  \hline 
  \\[-2mm]
   $N$            & $\phantom{-}117\pm16\phantom{0}$ 
   \\     
   $\updelta m_{\mathrm{BW}}$   
   &  $-273\pm61\phantom{0}\kevcc$   
   \\
   $\Gamma_{\mathrm{BW}}$ 
    & $\phantom{-}410\pm165\kev$ 

	\end{tabular*}
	\vspace{3mm}
\end{table}

To~validate the~presence of the~signal component, 
several additional cross\nobreakdash-checks are
performed.
The~data are categorised according 
to data\nobreakdash-taking 
periods including the~polarity of the~LHCb
dipole magnet and the~charge 
of the~\Tcc~candidates.
%
Instead of statistically subtracting
the~non\nobreakdash-\Dz~background,
the~mass of 
each \mbox{$\decay{\Dz}{\Km\pip}$}~candidate 
is required to be within a narrow~region around 
the~known mass of the~\Dz~meson~\cite{PDG2021}.
The~results are found to be consistent
among all samples and analysis techniques.
Furthermore, dedicated studies are performed 
to ensure that the observed signal  is not caused by
kaon or pion misidentification, 
doubly Cabibbo\nobreakdash-suppressed 
\mbox{$\decay{\Dz}{\Kp\pim}$}~decays
and $\Dz\Dzb$~oscillations,
decays of charm hadrons originating 
from beauty hadrons, 
or artefacts due to the~track reconstruction
creating duplicate tracks.

Systematic uncertainties 
for the~$\updelta m_{\mathrm{BW}}$
and 
$\Gamma_{\mathrm{BW}}$~parameters 
are summarised in Table~\ref{tab:systematic}
and described below. %
The~largest systematic uncertainty 
is related to the~fit model 
and is studied using pseudoexperiments 
with alternative parametrisations 
of the~\Dz\Dz\pip~mass shape.
Several
variations in 
the~fit model are considered: 
changes in the~signal model due 
to the~imperfect knowledge of the~detector resolution, 
an~uncertainty in 
the~correction factor for 
the~resolution 
taken from control channels,
parametrisation of the~background component
and the~additional model parameters 
of the~Breit\nobreakdash--Wigner function.
The~model uncertainty related to 
the~assumption of $\mathrm{J^P}=1^+$~quantum 
numbers of the~state 
is estimated and listed separately.
The~results are affected 
by the~overall detector momentum scale, 
which is known to 
a~relative precision of 
$\updelta\upalpha=
3\times10^{-4}$~\cite{LHCb-PAPER-2013-011}. 
The corresponding uncertainty is 
estimated using simulated samples where 
the~momentum-scale is modified by 
factors of $\left(1\pm\updelta\upalpha\right)$.
In~the reconstruction, 
the~momenta of charged tracks 
are corrected for energy loss 
in the~detector material,  
the amount of which is known 
with a~relative uncertainty
of 10. 
The~resulting uncertainty 
is assessed by varying 
the~energy loss correction 
by $\pm10\%$.
As the~mass of the~$\Dz\Dz\pip$~combinations is 
calculated with the mass of each $\Dz$~meson 
constrained to the~known value of the~\Dz~mass,
the~$\updelta m_{\mathrm{BW}}$~parameter
is insensitive to the~precision of the~\Dz~mass.
However, the~small uncertainty of 2\kevcc for
the $\Dstarp-\Dz$~mass difference~\cite{PDG2021}
directly affects 
the~$\updelta m_{\mathrm{BW}}$~value.
The~corresponding systematic uncertainty is added.

\ifthenelse{\boolean{wordcount}}{}{
\begin{table}[tb]
	\centering
	\caption{\small 
	Systematic uncertainties for 
	the~$\updelta m_{\mathrm{BW}}$ 
	and $\Gamma_{\mathrm{BW}}$~parameters.  
	The~total uncertainty is calculated 
	as the~sum in quadrature of all components 
	except for those related to 
	the~$\mathrm{J^P}$~quantum numbers
	assignment, which are handled separately. 
	}
	\label{tab:systematic}
	\vspace{2mm}
	\begin{tabular*}{0.75\textwidth}{@{\hspace{3mm}}l@{\extracolsep{\fill}}cc@{\hspace{3mm}}}
	Source 
    & $\upsigma_{\updelta
    m_{\mathrm{BW}}}~\left[\!\kevcc\right]$
    & $\upsigma_{\Gamma_{\mathrm{BW}}}~\left[\!\kev\right]$
   \\[1.5mm]
  \hline 
  \\[-2mm]
  Fit model                         &             &     \\
  ~~Resolution model                & 2   &  \phantom{0}7 \\
  ~~Resolution correction factor    & 1   &  30 \\
  ~~Background model                & 3   &  30 \\ 
  ~~Model parameters                & $<1\phantom{0}\,\,$ & $<1\phantom{0}$ \\ 
  Momentum scale                 &  3 &  ---   \\
  Energy loss corrections        &  1 &  ---  \\
  $\Dstarp-\Dz$ mass difference &  2 & --- 
  \\[1.5mm]
  \hline 
  \\[-2mm]
  Total  & 5 & 43
  \\[1.5mm]
  \hline
  \\[-2mm]
  $\mathrm{J^P}$~quantum numbers     & ${}^{+11}_{-14}$ & ${}^{+18}_{-38}$
   \end{tabular*}
	\vspace{3mm}
\end{table}
}

In summary,  using the~full
dataset corresponding to an~integrated 
luminosity 
of $9\invfb$,
collected by the~\lhcb experiment 
at centre-of-mass energies of 7, 8 and~13\tev
, 
a~narrow peak is observed 
in the~mass spectrum of $\Dz\Dz\pip$~candidates 
produced promptly in $\proton\proton$~collisions.
The~statistical significance of the peak 
is overwhelming. 
Using the~Breit\nobreakdash--Wigner 
parametrisation, 
the~location of the peak 
relative to the~$\Dstarp\Dz$~mass threshold,
$\updelta m_{\mathrm{BW}}$,
and the~width, $\Gamma_{\mathrm{BW}}$, 
are determined to be 
\ifthenelse{\boolean{wordcount}}{}{
\begin{eqnarray*}
\updelta m_{\mathrm{BW}} 
& = &  
-273\pm \phantom{0}61 
\pm \phantom{0}5\,  {}^{\,+\,11}_{\,-\,14} \kevcc \,,
\\
\Gamma_{\mathrm{BW}} 
& = & 
\phantom{-} 410 \pm 165 \pm  43\,   {}^{\,+\,18}_{\,-\,38}\kev\,,
\end{eqnarray*}
}
where the~first uncertainty is statistical, 
the~second systematic 
and the~third is related 
to the~$\mathrm{J^P}$~quantum numbers  assignment.
The~measured $\updelta m_{\mathrm{BW}}$  
value corresponds 
to a~mass of approximately 3875\mev.
This~is the~narrowest exotic state 
observed to date~\cite{Brambilla:2019esw,
PDG2021}.
The~minimal quark content 
for this
state is $\cquark\cquark\uquarkbar\dquarkbar$.
Two~heavy quarks of the~same 
flavour make it manifestly exotic, 
\ie beyond the conventional pattern
of hadron formation found in mesons and baryons.
Moreover, 
a~combination of the~near\nobreakdash-threshold mass,
narrow decay width and 
its appearance in prompt hadroproduction
demonstrates its genuine resonance nature.
The~measured mass and width are consistent with 
the~expected values for a~\Tcc~isoscalar 
tetraquark ground state with 
quantum numbers $\mathrm{J}^{\mathrm{P}}=1^+$. %
The~precision of the~mass measurement 
with respect to the~corresponding threshold 
is superior to those of all other exotic states, 
which will give better understanding of 
the~nature of exotic states. %
A~dedicated study of the~reaction amplitudes
for the~\mbox{$\decay{\Tcc}{\Dz\Dz\pip}$} and
\mbox{$\decay{\Tcc}{\Dz\Dp\piz(\g)}$}~decays
that uses the~isospin 
symmetry for~the~\mbox{$\decay{\Tcc}{\Dstar\D}$}
transition~\cite{LHCb-PAPER-2021-032}
yields insights on
the~fundamental resonance properties, 
like the~pole position, 
the scattering length and the~effective range.
The~observation of 
this $\cquark\cquark\uquarkbar\dquarkbar$~tetraquark 
candidate 
close to the~$\Dstarp\Dz$ threshold 
provides~strong support 
for the~theory approaches and models 
that predict 
the~existence 
of a~$\bquark\bquark\uquarkbar\dquarkbar$~tetraquark 
stable with respect to the~strong and 
electromagnetic interactions.
%

\clearpage
\section*{Methods}

\subsection*{Experimental setup} 


The \lhcb detector~\cite{LHCb-DP-2008-001,LHCb-DP-2014-002} is a single-arm forward
spectrometer covering the~pseudorapidity range $2<\eta <5$,
designed for the study of particles containing \bquark or \cquark
quarks. The detector includes a high-precision tracking system
consisting of a~silicon\nobreakdash-strip vertex detector surrounding 
the~$\proton\proton$~interaction region,
a~large-area silicon-strip detector located
upstream of a~dipole magnet with a~bending power of about
$4{\mathrm{\,Tm}}$, and three stations of silicon\nobreakdash-strip detectors and straw
drift tubes
placed downstream of the magnet.
The~tracking system provides a measurement of the momentum, \ptot, 
of charged particles with
a~relative uncertainty that varies from 0.5\% at low momentum to 1.0\% at 200\gevc.
The minimum distance of a track to 
a~primary $\proton\proton$~collision vertex\,(PV), 
the~impact parameter\,(IP), 
is measured with a resolution of $(15+29/\pt)\mum$,
where \pt is the component of the momentum transverse to the beam, in\,\gevc.
Different types of charged hadrons are distinguished using information
from two ring-imaging Cherenkov detectors~\cite{LHCb-DP-2012-003}. 
Photons, electrons and hadrons are identified by 
a~calorimeter system consisting of
scintillating-pad and preshower detectors, an electromagnetic
and a hadronic calorimeter. 
Muons are identified by a~system composed of 
alternating layers of iron and multiwire
proportional chambers.
The~online event selection is performed by 
a~trigger, 
which consists of a~hardware stage, based on information from 
the~calorimeter and muon systems, 
followed by a~software stage, which applies a full event
reconstruction.
The~trigger selection algorithms are primarily 
based on identifying key characteristics 
of 
beauty and charm hadrons
and their decay products, 
such as high \pt final state particles, 
and a~decay vertex that is significantly displaced 
from any of the~$\proton\proton$~interaction 
vertices in the~event.

\subsection*{Simulated samples} 

Simulation is required to model 
the~effects of the detector acceptance,
resolution and the~imposed selection requirements.
In~the~simulation, 
$\proton\proton$~collisions are generated using
  \pythia 
  with a~specific \lhcb configuration~\cite{LHCb-PROC-2010-056}.
  Decays of unstable particles
  are described by \evtgen~\cite{Lange:2001uf}.
  The~interaction of the~generated particles with the detector, and its response,
  are implemented using the~\geant
  toolkit 
  as described in
  Ref.~\cite{LHCb-PROC-2011-006}. 

\subsection*{Selection}
The~selection 
of \mbox{$\decay{\Dz}{\Km\pip}$}~candidates
and \mbox{$\Dz\Dz\pip$}~combinations 
is similar to those used in 
Ref.~\cite{LHCb-PAPER-2012-003}.
Kaon and  pion candidates  are selected from 
well\nobreakdash-reconstructed tracks
within the~acceptance of the~spectrometer.
Particle identification 
is provided using information from
the~ring\nobreakdash-imaging 
Cherenkov detectors.
Kaons and pions that have transverse 
momenta larger than $250\mevc$ 
and are inconsistent with being produced 
in a~$\proton\proton$~interaction 
vertex are combined together to form $\Dz$~candidates.
The~resulting $\Dz$~candidates are 
required to have 
good vertex quality,
mass within $\pm 65\mevcc$ 
of the~known \Dz~mass~\cite{PDG2021}
(mass resolution for the~\mbox{$\decay{\Dz}{\Km\pip}$}
signal is~7\mevcc), 
transverse momentum larger than $1\gevc$, 
decay time larger than $100\mum/c$ and 
a~momentum  direction that is consistent 
with the~vector from the~primary to 
the~secondary vertex. 
Selected pairs of \Dz~candidates consistent 
with originating from a~common primary vertex 
are then combined with 
pion candidates of the~same  charge 
as the~pions from 
the~\mbox{$\decay{\Dz}{\Km\pip}$}~decay
candidates  to 
form $\Dz\Dz\pip$~candidates. 
At~least one of the two $\Dz\pip$ 
combinations is required to 
have good vertex quality and mass 
not exceeding the~known 
$\Dz$~mass by more than $155\mevcc$.
For~each~$\Dz\Dz\pip$ candidate 
a~kinematic fit~\cite{Hulsbergen:2005pu} 
is performed.
This~fit 
requires both $\Dz$~candidates and a~pion  
to originate from the~same primary vertex.
A~requirement on the~quality
of this fit is applied to further 
suppress combinatorial background 
and reduce the background from $\Dz$~candidates 
produced in two independent 
$\proton\proton$~interactions 
or in decays of 
beauty hadrons~\cite{LHCb-PAPER-2012-003}.
To~suppress background from kaon 
and pion candidates reconstructed from 
one common track,
all track pairs of the~same charge are required to have 
opening angle inconsistent with being zero 
and mass of the~combination to be inconsistent 
with the~sum of masses of the~two constituents.

\subsection*{Non-\Dz background subtraction}

The~two\nobreakdash-dimensional 
distribution of the~mass of 
one \Dz~candidate versus
the~mass of the~other \Dz~candidate from 
selected  $\Dz\Dz\pip$~combinations 
is used to subtract the~background 
from fake \Dz~candidates. 
The~procedure employs 
the~\sPlot technique~\cite{Pivk:2004ty},
where an~extended unbinned 
maximum\nobreakdash-likelihood
fit to the~two\nobreakdash-dimensional 
distribution is performed. 
The~signal is described  
using a~modified 
Novosibirsk function 
and the~background is modelled
by a~product of 
an~exponential function 
and a~positive polynomial 
function~\cite{LHCb-PAPER-2012-003}. 
Each candidate 
is assigned 
a~positive weight 
for being signal\nobreakdash-like or
a~negative weight for 
being
background\nobreakdash-like, 
with the~masses 
of the~two \Dz~candidates 
as the~discriminating 
variables. 
All~candidates are then retained 
and the~weights are used in 
the~further analysis
for statistical subtraction of 
non\nobreakdash-\Dz~background.

\subsection*{Contributions from the~$\Dz\Dzb$~oscillations}
A~hypothetical narrow 
charmonium\nobreakdash-like  
state decaying into 
the~$\Dz\Dzb\pip$~final state, 
followed by the~$\Dzb\to\Dz$~transition 
or doubly Cabibbo\nobreakdash-suppressed decay
\mbox{$\decay{\Dzb}{\Km\pip}$}, 
would produce a~narrow signal in 
the~reconstructed $\Dz\Dz\pip$~mass spectrum. 
If the~observed narrow 
near\nobreakdash-threshold peak in 
the~reconstructed 
\Dz\Dz\pip system  is caused  
by the~$\Dz\Dzb$~oscillations or 
doubly Cabibbo\nobreakdash-suppressed 
decays, a~much larger signal should be 
visible in the~reconstructed $\Dz\Dzb\pip$~mass spectrum 
at the~same mass.
No such structure 
is observed,
see Fig.~9 in Ref.~\cite{LHCb-PAPER-2021-032}.

\subsection*{Systematic uncertainties} 

Several sources of systematic uncertainty 
on the~mass $\updelta m_{\mathrm{BW}}$
and width $\Gamma_{\mathrm{BW}}$
of the~\Tcc~state 
have been evaluated. 
The~largest systematic uncertainty 
is related to the~fit model 
and is studied using a~set of alternative
parametrisations and pseudoexperiments.
For~each alternative model, 
an~ensemble of pseudoexperiments is produced;
each is generated using the~model under 
consideration
with parameters obtained from a~fit~to 
the~data.  
A~subsequent~fit with the~default
model to 
each pseudoexperiment is performed 
and the~mean values of 
the~parameters of interest over 
the~ensemble are evaluated.
The~absolute value of the~difference between 
the~ensemble mean 
and 
the~value of the~parameter obtained from 
the~fit to the~data sample is used to 
characterise the~difference
between the~alternative  model 
and the~default model. 
The~maximal value of such a difference 
over the~considered set of 
alternative models is taken 
as the~corresponding systematic uncertainty
for the~mass $\updelta m_{\mathrm{BW}}$
and width $\Gamma_{\mathrm{BW}}$
of the~\Tcc~state.
The~following sources of systematic uncertainties 
related to the~fit model are considered: 
\begin{itemize} 
\item Imperfect knowledge of detector resolution model: 
to~estimate the~associated systematic uncertainty 
alternative resolution functions are studied, namely 
a~symmetric variant 
of an~Apollonios function 
a~modified Gaussian function with 
symmetric power\nobreakdash-law 
tails on  both sides of 
the~distribution 
a~generalised symmetric
Student's $t$\nobreakdash-distribution; 
a~symmetric Johnson's 
$\mathrm{S_U}$~distribution 
and a~modified Novosibirsk function. 
\item Difference in detector 
resolution due to imperfect modelling:
a~correction~factor of 1.05 for the~resolution 
is applied for the~default fit to 
account for such a~difference. 
This factor was studied for 
several other decays measured with 
the~LHCb detector
and found 
to lie between~1.0 and 
1.1~\cite{LHCb-PAPER-2020-008,LHCb-PAPER-2020-009}
For~decays with relatively 
low\nobreakdash-momentum tracks, 
this factor is close to~1.05. 
The~factor is also 
cross\nobreakdash-checked using large samples of 
\mbox{$\decay{\Dstarp}{\Dz\pip}$}~decays,
where a~value of~1.06 is obtained.
To~assess the~systematic uncertainty 
related to this factor,
detector resolution models with correction 
factors of 1.0 and 1.1 are studied 
as alternative models. 

\item Parametrisation of the~background component: 
to~assess the~associated systematic uncertainty, 
the~order of the~positive polynomial function 
used for the~baseline fit is 
varied. In~addition, to estimate the~possible 
effect of a~small contribution 
from $\Dz\Dz\pip$~combinations without 
an~intermediate \Dstarp~meson,
a~three\nobreakdash-body background component 
is added to the~fit.
This~component is described by a~product 
of the~three\nobreakdash-body phase space 
function  
and a~positive linear or 
second\nobreakdash-order polynomial function. 
The~contribution from 
non\nobreakdash-resonant 
$\Dz\Dz\pip$~background 
is negligible 
in the~low\nobreakdash-mass 
region due to the~$O(Q^2)$~scaling of 
the~three\nobreakdash-body 
phase\nobreakdash-space factor near threshold.

\item Model parameters 
of the~Breit\nobreakdash--Wigner function: 
alternative parametrisations include 
different choices 
for the~decay structure, 
$m_{\PA}=m_{\Dz}$ and $m_{\PB}=m_{\Dz}+m_{\pip}$; 
the~meson 
radius, $1.5\gev^{-1}$ and $5\gev^{-1}$,
and the~orbital angular momentum 
between $\PA$ 
and $\PB$~particles, 
corresponding to S-~and D\nobreakdash-waves.
The~effect of the~different decay structure and 
choice of meson radius
is smaller than 1\kevcc and 1\kev
for the~$\updelta m_{\mathrm{BW}}$
and $\Gamma_{\mathrm{BW}}$~parameters, respectively. 
The~parameters of interest are more sensitive to the~choice
of orbital angular momentum, 
in which the~S\nobreakdash-wave 
function gives larger $\updelta m_{\mathrm{BW}}$
and smaller $\Gamma_{\mathrm{BW}}$, 
and the~D\nobreakdash-wave function 
corresponds to 
smaller $\updelta m_{\mathrm{BW}}$
and larger $\Gamma_{\mathrm{BW}}$.
%
As~the~S\nobreakdash-wave and 
D\nobreakdash-wave imply 
that the~quantum numbers of 
the~\Tcc state differ from $\mathrm{J^P=1^+}$,
the~corresponding systematic 
uncertainty is considered separately 
and is not included  
it in the~total systematic uncertainty.

\end{itemize} 

The~calibration of the momentum scale of 
the~tracking system 
is based upon large calibration 
samples of  \mbox{$\decay{\Bu}{\jpsi\Kp}$} 
and  
\mbox{$\decay{\jpsi}
{\mumu}$}~decays. 
The~accuracy of the~procedure has been checked 
using other fully reconstructed $\B$~decays together 
with two\nobreakdash-body
$\PUpsilon(\mathrm{nS})$ and 
$\KS$~decays and 
the~largest deviation of the~bias 
in the~momentum  scale
of $\updelta\upalpha=3\times10^{-4}$~is 
taken as the~uncertainty~\cite{LHCb-PAPER-2013-011}.
This is then propagated to uncertainties for 
the~parameters of  interest using simulated samples,
where momentum scale corrections 
of $\left(1\pm\updelta\upalpha\right)$
are applied. Half of the~difference 
between the~peak locations obtained 
with $1+\updelta\upalpha$ 
and $1-\updelta\upalpha$~corrections 
applied to 
the~same simulated sample
is taken as an~estimate of the~systematic 
uncertainty due 
to the~momentum scale. 
The~main contribution to this uncertainty 
is due to the~bachelor pion track, 
since  the~\Dz~mass~constraint
reduces the~contributions from 
the~kaon and pion tracks originating 
from \Dz~meson decays.

In~the~reconstruction the~momenta 
of the~charged tracks 
are corrected for the~energy loss in the~detector 
material. 
The~energy loss corrections are calculated  
using  the~Bethe\nobreakdash--Bloch
formula. 
The~amount of material traversed in the~tracking
system by a~charged particle 
is known to~10\% accuracy.  
To~assess the~corresponding uncertainty, 
the~magnitude of the~calculated corrections 
is varied by $\pm10\%$.
Half of the~difference 
between the~peak locations obtained with $+10\%$ 
and $-10\%$~corrections applied to the~same simulated sample
is taken as an~estimate of the~systematic 
uncertainty due to the~energy loss corrections.

The~mass of the $\Dz\Dz\pip$~combinations is 
calculated 
with both \Dz~candidate masses 
constrained 
to the~known $\Dz$~meson mass~\cite{PDG2021}.
This~procedure removes 
the~uncertainty on the~$\updelta m_{\mathrm{BW}}$ 
parameter related to 
imprecise knowledge of the~\Dz~mass.
In~contrast, the~small uncertainty of 2\kevcc for 
the~known 
$\Dstarp-\Dz$~mass difference~\cite{
  PDG2021}
directly affects the~$\updelta m_{\mathrm{BW}}$ value
and therefore is assigned as 
the~corresponding systematic uncertainty.

\subsection*{Acknowledgements}
%
%
This Letter is dedicated 
to the~memory of our dear colleague 
Simon Eidelman, whose friendship, 
deep physics insights and contributions 
to improving the~quality of our papers 
were greatly appreciated and will be missed. 
We~express our gratitude to our colleagues in the~CERN
accelerator departments for the~excellent performance of the~LHC. 
We~thank the technical and administrative staff at the LHCb
institutes.
We acknowledge support from CERN 
and from the national agencies:
CAPES, CNPq, FAPERJ and FINEP\,(Brazil); 
MOST and NSFC\,(China); 
CNRS/IN2P3\,(France); 
BMBF, DFG and MPG\,(Germany); 
INFN\,(Italy); 
NWO\,(Netherlands); 
MNiSW and NCN\,(Poland); 
MEN/IFA\,(Romania); 
MSHE\,(Russia); 
MICINN\,(Spain); 
SNSF and SER\,(Switzerland); 
NASU\,(Ukraine); 
STFC\,(United Kingdom); 
DOE NP and NSF\,(USA).
We~acknowledge the~computing resources that are provided by CERN, 
IN2P3\,(France), 
KIT and DESY\,(Germany), 
INFN\,(Italy), 
SURF\,(Netherlands),
PIC\,(Spain), 
GridPP\,(United Kingdom), 
RRCKI and Yandex~LLC\,(Russia), 
CSCS\,(Switzerland), 
IFIN-HH\,(Romania), 
CBPF\,(Brazil),
PL\nobreakdash-GRID\,(Poland) and NERSC\,(USA).
We are indebted to the~communities behind 
the~multiple open-source
software packages on which we depend.
Individual groups or members have received support from
ARC and ARDC\,(Australia);
AvH Foundation\,(Germany);
EPLANET, Marie Sk\l{}odowska-Curie Actions and ERC\,(European Union);
A*MIDEX, ANR, IPhU and Labex P2IO, and R\'{e}gion 
Auvergne-Rh\^{o}ne-Alpes\,(France);
Key Research Program of Frontier Sciences of CAS, 
CAS~PIFI, CAS~CCEPP, 
Fundamental Research Funds for the~Central Universities, 
and Sci. \& Tech. Program of Guangzhou\,(China);
RFBR, RSF and Yandex LLC\,(Russia);
GVA, XuntaGal and GENCAT\,(Spain);
the~Leverhulme Trust, the~Royal Society
 and UKRI\,(United Kingdom).

\subsection*{Author Contribution Statement} 
All contributing authors, as listed at the~end of
this manuscript, 
have contributed to the~publication, 
being variously involved in the~design 
and the~construction of the~detector, 
in writing software, calibrating sub-systems, 
operating the~detector and acquiring data 
and finally analysing the~processed data.

\subsection*{Competing Interests Statement}
The authors declare no competing interests.

\subsection*{Correspondence and requests for materials}
Correspondence and requests for materials
should be addressed to I.~Belyaev \href{mailto:Ivan.Belyaev@itep.ru}{Ivan.Belyaev@itep.ru}.

\subsection*{Data Availability Statement} 

LHCb data used in this analysis will be released according to the LHCb external data access
policy, that can be downloaded from 
\href{http://opendata.cern.ch/record/410/files/LHCb-Data-Policy.pdf}
    {\tt{http://opendata.cern.ch/record/410/files/LHCb-Data-Policy.pdf}}.
    
The raw data in all of the figures of this manuscript
can be downloaded from 
\href{https://cds.cern.ch/record/2780001}{\tt{https://cds.cern.ch/record/2780001}},
where no access codes are required. In addition, 
the~unbinned background\nobreakdash-subtracted data, 
shown in Fig.~\ref{fig:DATA_BW0_fit}
have been added to the~{\sc{HEPData}} record at
\href{https://www.hepdata.net/record/ins1915457} 
{\tt{https://www.hepdata.net/record/ins1915457}}.

\subsection*{Code Availability Statement} 
LHCb software used to process the~data analysed in this manuscript 
is available at {\sc{GitLab}} repository 
\href{https://gitlab.cern.ch/lhcb}{\tt{https://gitlab.cern.ch/lhcb}}.
The~specific software used in data analysis 
is available at {\sc{Zenodo}} repository
\href{https://zenodo.org/record/5595937}{\tt{DOI:10.5281/zenodo.5595937}}.

\clearpage
\addcontentsline{toc}{section}{References}
\setboolean{inbibliography}{true}
\bibliographystyle{LHCb}
\bibliography{main,standard,LHCb-PAPER,LHCb-CONF,LHCb-DP,LHCb-TDR}

\clearpage
\centerline
{\large\bf LHCb collaboration}
\begin
{flushleft}
\small
R.~Aaij$^{1}$,
A.S.W.~Abdelmotteleb$^{2}$,
C.~Abell{\'a}n~Beteta$^{3}$,
F.J.~Abudinen~Gallego$^{2}$,
T.~Ackernley$^{4}$,
B.~Adeva$^{5}$,
M.~Adinolfi$^{6}$,
H.~Afsharnia$^{7}$,
C.~Agapopoulou$^{8}$,
C.A.~Aidala$^{9}$
% ...

{\footnotesize \it

$^{1}$Nikhef National Institute for Subatomic Physics, Amsterdam, Netherlands\\
$^{2}$Department of Physics, University of Warwick, Coventry, United Kingdom\\
$^{3}$Physik-Institut, Universit{\"a}t Z{\"u}rich, Z{\"u}rich, Switzerland\\
$^{4}$Oliver Lodge Laboratory, University of Liverpool, Liverpool, United Kingdom\\
$^{5}$Instituto Galego de F{\'\i}sica de Altas Enerx{\'\i}as (IGFAE), Universidade de Santiago de Compostela, Santiago de Compostela, Spain\\
$^{6}$H.H. Wills Physics Laboratory, University of Bristol, Bristol, United Kingdom\\
$^{7}$Universit{\'e} Clermont Auvergne, CNRS/IN2P3, LPC, Clermont-Ferrand, France\\
$^{8}$LPNHE, Sorbonne Universit{\'e}, Paris Diderot Sorbonne Paris Cit{\'e}, CNRS/IN2P3, Paris, France\\
$^{9}$University of Michigan, Ann Arbor, United States, associated to $^{25}$\\
% ...
$^{25}$Syracuse University, Syracuse, NY, United States\\
% ...
}
\end{flushleft}

\end{document}